\newcommand\apjcls{1}
\newcommand\aastexcls{2}
\newcommand\othercls{3}

\newcommand\papercls{\aastexcls}
\documentclass[tighten, times,twocolumn]{aastex62}  

\if\papercls \apjcls
\usepackage{apjfonts}
\else\if\papercls \othercls
\usepackage{epsfig}
\usepackage{margin}
\usepackage{times}
\fi\fi
\usepackage{ifthen}
\usepackage{natbib}
\usepackage{bm}
\usepackage{amssymb, amsmath}
\usepackage{appendix}
\usepackage{etoolbox}
\usepackage[T1]{fontenc}
\usepackage{paralist}
\usepackage{newtxtext,newtxmath}
\usepackage{soul}
\if\papercls \apjcls 
\newcommand\aas{\ref@jnl{AAS Meeting Abstracts}}
\newcommand\dps{\ref@jnl{AAS/DPS Meeting Abstracts}}
\newcommand\maps{\ref@jnl{MAPS}}
\else\if\papercls \othercls
\usepackage{astjnlabbrev-jh}
\fi\fi

\if\papercls \aastexcls
\hypersetup{citecolor=blue, 
            linkcolor=blue, 
            menucolor=blue, 
            urlcolor=blue}  
\else
\usepackage[
bookmarks=true,           
bookmarksnumbered=true,   
colorlinks=true,          
citecolor=blue,           
linkcolor=blue,           
menucolor=blue,           
urlcolor=blue,            
linkbordercolor={0 0 1},  
pdfborder={0 0 1},
frenchlinks=true]{hyperref}
\fi

\if\papercls \othercls

\else

\fi

\providecommand{\adsurl}[1]{\href{#1}{ADS}}

\makeatletter
\patchcmd{\NAT@citex}
  {\@citea\NAT@hyper@{%
     \NAT@nmfmt{\NAT@nm}%
     \hyper@natlinkbreak{\NAT@aysep\NAT@spacechar}{\@citeb\@extra@b@citeb}%
     \NAT@date}}
  {\@citea\NAT@nmfmt{\NAT@nm}%
   \NAT@aysep\NAT@spacechar\NAT@hyper@{\NAT@date}}{}{}

\patchcmd{\NAT@citex}
  {\@citea\NAT@hyper@{%
     \NAT@nmfmt{\NAT@nm}%
     \hyper@natlinkbreak{\NAT@spacechar\NAT@@open\if*#1*\else#1\NAT@spacechar\fi}%
       {\@citeb\@extra@b@citeb}%
     \NAT@date}}
  {\@citea\NAT@nmfmt{\NAT@nm}%
   \NAT@spacechar\NAT@@open\if*#1*\else#1\NAT@spacechar\fi\NAT@hyper@{\NAT@date}}
  {}{}
\makeatother

\makeatletter
\DeclareRobustCommand{\lowcase}[1]{\@lowcase#1\@nil}
\def\@lowcase#1\@nil{\if\relax#1\relax\else\MakeLowercase{#1}\fi}
\makeatother

\DeclareSymbolFont{UPM}{U}{eur}{m}{n}
\DeclareMathSymbol{\umu}{0}{UPM}{"16}
\let\oldumu=\umu
\renewcommand\umu{\ifmmode\oldumu\else\math{\oldumu}\fi}

\if\papercls \othercls

\else

\fi

\let\oldsim=\sim
\renewcommand\sim{\ifmmode\oldsim\else\math{\oldsim}\fi}
\let\oldpm=\pm
\renewcommand\pm{\ifmmode\oldpm\else\math{\oldpm}\fi}
\newcommand\by{\ifmmode\times\else\math{\times}\fi}

\newbox{\wdbox}
\renewcommand\c{\setbox\wdbox=\hbox{,}\hspace{\wd\wdbox}}
\renewcommand\i{\setbox\wdbox=\hbox{i}\hspace{\wd\wdbox}}

\newcount\timect
\newcount\hourct
\newcount\minct
\newcommand\now{\timect=\time \divide\timect by 60
         \hourct=\timect Cltiply\hourct by 60
         \minct=\time \advance\minct by -\hourct
         \number\timect:\ifnum \minct < 10 0\fi\number\minct}

\catcode`@=11

\newcommand\comment[1]{}

\newcommand\commenton{\catcode`\%=14}

\renewcommand\math[1]{$#1$}
\newcommand\mathshifton{\catcode`\$=3}

\let\atab=&
\newcommand\atabon{\catcode`\&=4}

\let\oldmsp=\sp
\let\oldmsb=\sb
\def\sp#1{\ifmmode
           \oldmsp{#1}%
         \else\strut\raise.85ex\hbox{\scriptsize #1}\fi}
\def\sb#1{\ifmmode
           \oldmsb{#1}%
         \else\strut\raise-.54ex\hbox{\scriptsize #1}\fi}
\newbox\@sp
\newbox\@sb
\def\sbp#1#2{\ifmmode%
           \oldmsb{#1}\oldmsp{#2}%
         \else
           \setbox\@sb=\hbox{\sb{#1}}%
           \setbox\@sp=\hbox{\sp{#2}}%
           \rlap{\copy\@sb}\copy\@sp
           \ifdim \wd\@sb >\wd\@sp
             \hskip -\wd\@sp \hskip \wd\@sb
           \fi
        \fi}
\def\msp#1{\ifmmode
           \oldmsp{#1}
         \else \math{\oldmsp{#1}}\fi}
\def\msb#1{\ifmmode
           \oldmsb{#1}
         \else \math{\oldmsb{#1}}\fi}

\def\supon{\catcode`\^=7}

\def\subon{\catcode`\_=8}

\def\supsubon{\supon \subon}

\newcommand\actcharon{\catcode`\~=13}

\newcommand\paramon{\catcode`\#=6}

\comment{And now to turn us totally on and off...}

\newcommand\reservedcharson{ \commenton  \mathshifton  \atabon  \supsubon 
                             \actcharon  \paramon}

\catcode`@=12
\reservedcharson

\if\papercls \apjcls

\else

\fi

\newcommand\chisq{\ifmmode{\chi\sp{2}}\else\math{\chi\sp{2}}\fi}
\newcommand\redchisq{\ifmmode{ \chi\sp{2}\sb{\rm red}}
                    \else\math{\chi\sp{2}\sb{\rm red}}\fi}
\newcommand\Teq{\ifmmode{T\sb{\rm eq}}\else$T$\sb{eq}\fi}
\newcommand\mjup{\ifmmode{M\sb{\rm Jup}}\else$M$\sb{Jup}\fi}
\newcommand\rjup{\ifmmode{R\sb{\rm Jup}}\else$R$\sb{Jup}\fi}
\newcommand\msun{\ifmmode{M\sb{\odot}}\else$M\sb{\odot}$\fi}
\newcommand\rsun{\ifmmode{R\sb{\odot}}\else$R\sb{\odot}$\fi}
\newcommand\mearth{\ifmmode{M\sb{\oplus}}\else$M\sb{\oplus}$\fi}
\newcommand\rearth{\ifmmode{R\sb{\oplus}}\else$R\sb{\oplus}$\fi}

\shorttitle{A Short Intense Dynamo at the Onset of Crystallization in White Dwarfs}
\shortauthors{Fuentes {\em et al.}}

\begin{document}

\title{A Short Intense Dynamo at the Onset of Crystallization in White Dwarfs}

\author{J. R. Fuentes}
\affiliation{\rm Department of Applied Mathematics, University of Colorado Boulder, Boulder, CO 80309-0526, USA}

\author{Matias Castro-Tapia}
\affiliation{\rm Department of Physics and Trottier Space Institute, McGill University, Montreal, QC H3A 2T8, Canada}

\author{Andrew Cumming}
\affiliation{\rm Department of Physics and Trottier Space Institute, McGill University, Montreal, QC H3A 2T8, Canada}

\begin{abstract}
The origin of large magnetic fields ($\gtrsim 10^6~\mathrm{G}$) in isolated white dwarfs is not clear. One possible explanation is that crystallization of the star's core drives compositional convection, which when combined with the star's rotation, can drive a dynamo. However, whether convection is efficient enough to explain the large intensity of the observed magnetic fields is still under debate. Recent work has shown that convection in cooling white dwarfs spans two regimes: efficient convection at the onset of crystallization, and thermohaline convection during most of the star's cooling history. Here, we calculate the properties of crystallization-driven convection for cooling models of several white dwarfs of different masses. We combine mixing-length theory with scalings from magneto-rotational convection to estimate the typical magnitude of the convective velocity and induced magnetic field for both scenarios. In the thermohaline regime, we find velocities $\sim 10^{-6}$--$10^{-5}~\mathrm{cm~s^{-1}}$, with fields restricted to $\lesssim~100~\mathrm{G}$. However, when convection is efficient, the flow velocity can reach  magnitudes of $\sim 10^2$--$10^3~\mathrm{cm~s^{-1}}$, with fields of $\sim 10^6$--$10^8~\mathrm{G}$, independent of the star's rotation rate. Thus, dynamos driven at the onset of crystallization could explain the large intensity magnetic fields measured for single white dwarfs.

\end{abstract}

\keywords{stars: interiors, stars: magnetic fields, white dwarfs}

\section{Introduction} 

One of the most striking phenomena in astrophysics is the ubiquity of magnetic fields. Like the Earth, stars and stellar remnants such as white dwarfs have a magnetic field. It is known that magnetic fields in a fraction of white dwarfs (WDs) can be more than a million times stronger than that of the Earth ($B_{\rm obs} \sim  10^4$--$10^9~\rm G$). However, the origin of such large magnetic fields has been a mystery \citep[see][for a review]{Ferrario2015,Ferrario2020}.

Recently, it has been suggested that white dwarfs create magnetic fields through a process similar to the dynamo in the Earth's core. This process involves the crystallization of the white dwarf's interior as it cools down \citep[e.g.,][]{kirzhnits1960,abrikosov1961,Salpeter1961,vanHorn1968}, which causes a separation of its chemical components \citep{Stevenson1980}. 
The oxygen within the white dwarf preferentially goes into the solid phase, with a corresponding amount of  carbon remaining in the liquid phase.
This separation results in compositional convection above the crystallized core due to the carbon's buoyancy \citep{ Schatzman1982,Mochkovitch1983, Isern1997}. With rotation added to the mix, convection can lead to a magnetic dynamo being sustained \citep{Isern2017,Belloni2021,Schreiber2021,Schreiber2022,Ginzburg2022}. 


Typically, the dynamo's strength is estimated by balancing the energy stored in the magnetic field with the kinetic energy carried by the convective motions $B^2/4\pi \sim \rho u^2_c$ (this is known as the \emph{equipartition or saturated dynamo hypothesis}), where $B$, $\rho$, and $u_c$ are the magnitude of the magnetic field, mass density, and convective velocity, respectively \citep{Christensen2009,Christensen2010}. To estimate $u_c$ during crystallization, previous studies \citep[e.g.,][]{Isern2017,Ginzburg2022} have assumed that the kinetic energy flux carried by the convective motions, $F_K$, is of the same order as the flux of gravitational energy released from chemical separation, $F_{\mathrm{grav}}$ (also $\sim$ convective heat flux, $F_H$). This leads to estimates for the convective velocities of $u_c\sim 10^2$--$10^6~\rm cm~s^{-1}$, which are enough to generate magnetic fields of $B \sim 10^{4}$--$10^{9}$ G, consistent with observations.

While some of the assumptions above may be correct, the large thermal conductivity of degenerate electrons in white dwarf interiors can reduce the efficiency of convection (i.e., convection occurs at small P\'eclet number, Pe, the ratio of thermal diffusion time to convective turnover time). In fact, previous estimations including thermal diffusion gave convective velocities of only $10^{-6}$--$10^{-1}~\rm cm~s^{-1}$, depending on the star's rotation rate \citep{Mochkovitch1983}. 

Recently, \cite{Fuentes2023} used mixing-length theory and CIA balance (a balance between the Coriolis, inertial, and buoyancy force; \citealt{Aurnou2020}) to investigate in detail the transport properties of compositionally-driven convection in white dwarfs. They found that thermal diffusion significantly reduces the convective speed to $ < 10^{-2}~\mathrm{cm~s^{-1}}$, in agreement with the early calculations by \cite{Mochkovitch1983}. Further, \cite{Fuentes2023} found that $F_{\mathrm{grav}} \sim F_H \gg F_K$, meaning that even though there is enough energy coming out from the star to explain the observed magnetic fields, only a small fraction is in the form of kinetic energy. When combined with the saturated dynamo hypothesis, those much smaller velocities result in magnetic fields of just a few G, in disagreement with observations. Those results have been recently confirmed by \cite{MontgomeryDunlap2023}, who used the transport prescriptions for thermohaline mixing in \cite{Brown2013} to study fluid mixing during phase separation in crystallizing white dwarfs.

The wide range of estimated velocities during crystallization-driven convection calls for a consistent theory to calculate the flow velocity and the resulting magnetic fields. We identify two ways to improve previous work. First, crystallization-driven convection spans two regimes during the evolution of cooling white dwarfs: A short-lived regime at the beginning of the crystallization, where the composition flux is large enough so that convection is efficient and $\mathrm{Pe}\gg 1$, and a second regime of inefficient convection (thermohaline convection) that lasts for most of the cooling history of the star \citep{CT24}. Second, the large intensity of observed magnetic fields suggest that CIA balance may not be at work in white dwarf interiors, and instead, MAC balance (a balance between the magnetic, buoyancy, and Coriolis force; \citealt{Stevenson1979}) controls the convective dynamics.

In this paper, we assess the plausibility of crystallization-driven dynamos by investigating how the MAC balance changes previous estimations for the convective velocities and the resulting magnetic field. In Section~
\ref{sec:cooling}, we discuss the cooling evolution for a canonical high-mass white dwarf. In Section~\ref{sec:dynamo}, we derive an analytic expression for the magnetic field induced by crystallization, as a function of different stellar parameters (including Pe). Then, in Section~\ref{sec:Pe} we estimate Pe for both regimes, small Pe or thermohaline convection (relevant for most of the white dwarf's evolution), and the regime of efficient convection or large Pe (important during the early phase of the crystallization).  In Section~\ref{sec:results} we show our main results, suggesting that dynamos have to be generated at the onset of crystallization. Finally, in Section~\ref{sec:discussion} we summarize and discuss our results. 

\section{Cooling history of massive white dwarfs}\label{sec:cooling}

In this section, we discuss the cooling evolution and crystallization of a $0.9 \msun$ white dwarf. We present a model of a slightly massive white dwarf because strong magnetic fields are mostly observed for WDs of masses $> 0.6\msun$ \citep{Ferrario2020}. Our model is one of those in \cite{CT24}, calculated with the MESA stellar evolution code version r23.05.1 \citep{Paxton2011, Paxton2013, Paxton2015, Paxton2018, Paxton2019, Jermyn2023} using a realistic composition profile from stellar evolution. Crystallization starts when the star reaches a luminosity $\sim 10^{-3}\ L_\sun$ corresponding to a central temperature $\approx 7\times 10^6\ {\rm K}$ at a central density of $1.7\times 10^6\ {\rm g\ cm^{-3}}$. The model follows the routine in \cite{Bauer2023} to account for the phase separation of the carbon-oxygen core, and uses the Skye EOS for dense matter in \cite{Jermyn2021}, which self-consistently determines the location of the liquid/solid phase transition and the associated latent heat release. More technical details of the model can be found in \cite{CT24}.

\begin{figure}
    \includegraphics[width=\columnwidth]{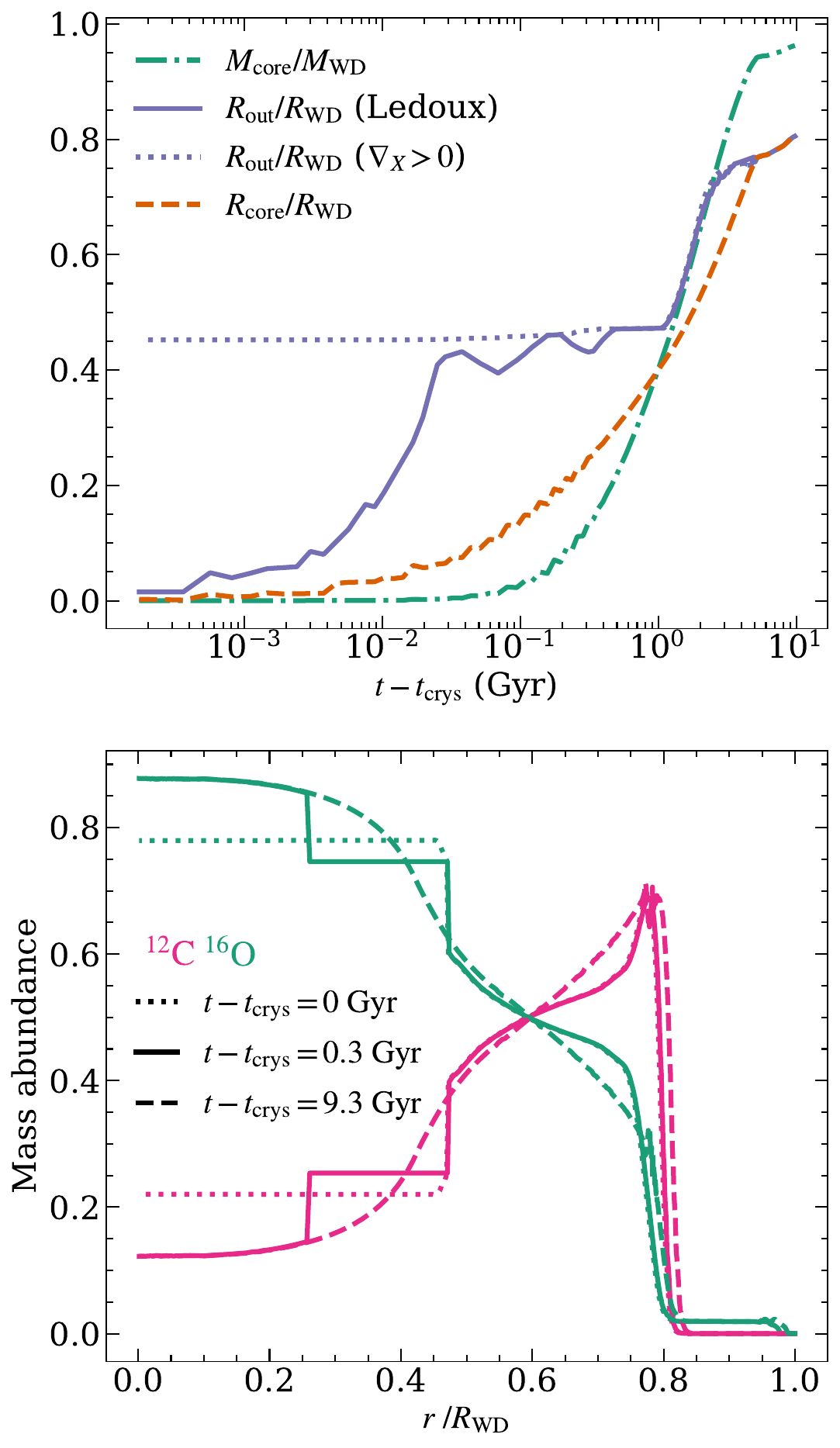}
    \caption{Top panel: Temporal evolution of the core mass, $M_{\rm core}$, core radius, $R_{\rm core}$, and outer radius of the convection zone, $R_{\rm out}$, once the white dwarf begins to crystallize. Results are shown for a cooling model of a $0.9\msun$ white dwarf obtained with the MESA stellar evolution code. For $R_{\mathrm{out}}$ we show two curves, corresponding to two different ways of determining the extent of the convection zone (the Ledoux criterion or just the unstable composition gradient ignoring the thermal buoyancy). Bottom panel: Chemical profiles of the C/O abundances at different times since the onset of crystallization for the model using the Ledoux criterion shown in the top panel. The profiles at $t-t_{\mathrm{crys}} = 0.3~\mathrm{Gyr}$ are shown during the regime of thermohaline convection (see text).}
    \label{fig:evolution}
\end{figure}

Figure~\ref{fig:evolution} (top panel) shows the evolution of the core radius and convection zone radius with time. As the star cools, the mass and radius of the core increase with time, while the outer boundary of the convective region moves outward at a different rate.
At the onset of crystallization, the convection zone spans $< 10\%$ of the star's radius. The C/O profiles at the beginning of crystallization have a well-mixed region that extends further out, up to $0.45R_{\mathrm{WD}}$ (see bottom panel). However, we find that thermal buoyancy stabilizes the unstable composition gradient through most of this region, leaving only a small inner part that is unstable to the Ledoux criterion ($\lesssim 0.1 R_{\mathrm{WD}}$) initially.
 
Subsequently, the convective boundary moves out to $\approx 0.45R_{\mathrm{WD}}$ on a timescale $\sim 1$--10~$\mathrm{Myr}$. This early evolution of the convection zone is not shown in the models of \cite{Isern2017} and \cite{Blatman2024} but the extent of the convection zone is similar to our models at $t-t_{\mathrm{crys}}> 0.1~\mathrm{Gyr}$. We suppose that this difference would be related to the time resolution of the data points shown in each case. At the same time, the rate at which the convection zone moves outwards in this phase depends on the amount of mixing at the convective boundary, and is therefore sensitive to the prescription used for convective boundary mixing. As a comparison, we also computed models in which the stabilizing effect of thermal buoyancy was ignored, finding that convection then quickly moves out to $0.45R_{\mathrm{WD}}$ (dotted curve in the top panel of Figure~\ref{fig:evolution}). The exact evolution of the convection zone in this early phase will require more careful modelling, likely multi-dimensional numerical simulations of convection.

At later times $\gtrsim 1\ \mathrm{Gyr}$, the convection zone shows a step-like behavior with time. This occurs because of the initial composition profile of the model. The jump in the carbon-oxygen abundances temporarily halts the convection zone until it can become carbon-rich enough to push out further (see bottom panel). As a result, the size of the convective region changes significantly along the cooling history of the star. After $\sim 5~\mathrm{Gyr}$, $R_{\rm out} \approx R_{\rm core}\approx 0.8R_{\rm WD}$ and the convection zone disappears. At this point, the core contains almost the entire mass of the star.

The efficiency of crystallization-driven convection is set by the strength of the mass flux of light elements (carbon) due to chemical separation at the solid-liquid interface, $F_X$. Since $F_X$ is controlled by the cooling rate of the core, the convective efficiency changes over the cooling history of white dwarfs. As showed by \citet{Fuentes2023} and more recently by \cite{CT24}, a natural dimensionless parameter that measures the strength of $F_X$ (and the efficiency) is
\begin{align} \label{eq:tau}
 \tau &= \left(\dfrac{F_X}{\rho H_P X}\right)\left(\dfrac{H_P^2}{\kappa_T}\right)\left(\dfrac{\chi_X}{\chi_T\nabla_{\mathrm{ad}}}\right)
=\left(\dfrac{t_{\mathrm{therm}}}{t_X}\right)\left(\dfrac{\chi_X}{\chi_T\nabla_{\mathrm{ad}}}\right)~,
\end{align}
where $\nabla_{\rm ad}$ is the adiabatic temperature gradient, $\chi_T = \left.\partial\ln P/\partial\ln T\right|_{\rho, X}$, $\chi_X = \left.\partial\ln P/\partial\ln X\right|_{\rho, T}$, $t_\mathrm{therm}=H_P^2/\kappa_T$ is the thermal diffusion time across a pressure scale height $H_P$ (where $\kappa_T$ is the thermal diffusivity), and $t_X=\rho H_P X/F_X$ is the timescale at which the light elements are released at the solid-liquid interface.

\begin{figure}
    \includegraphics[width=\columnwidth]{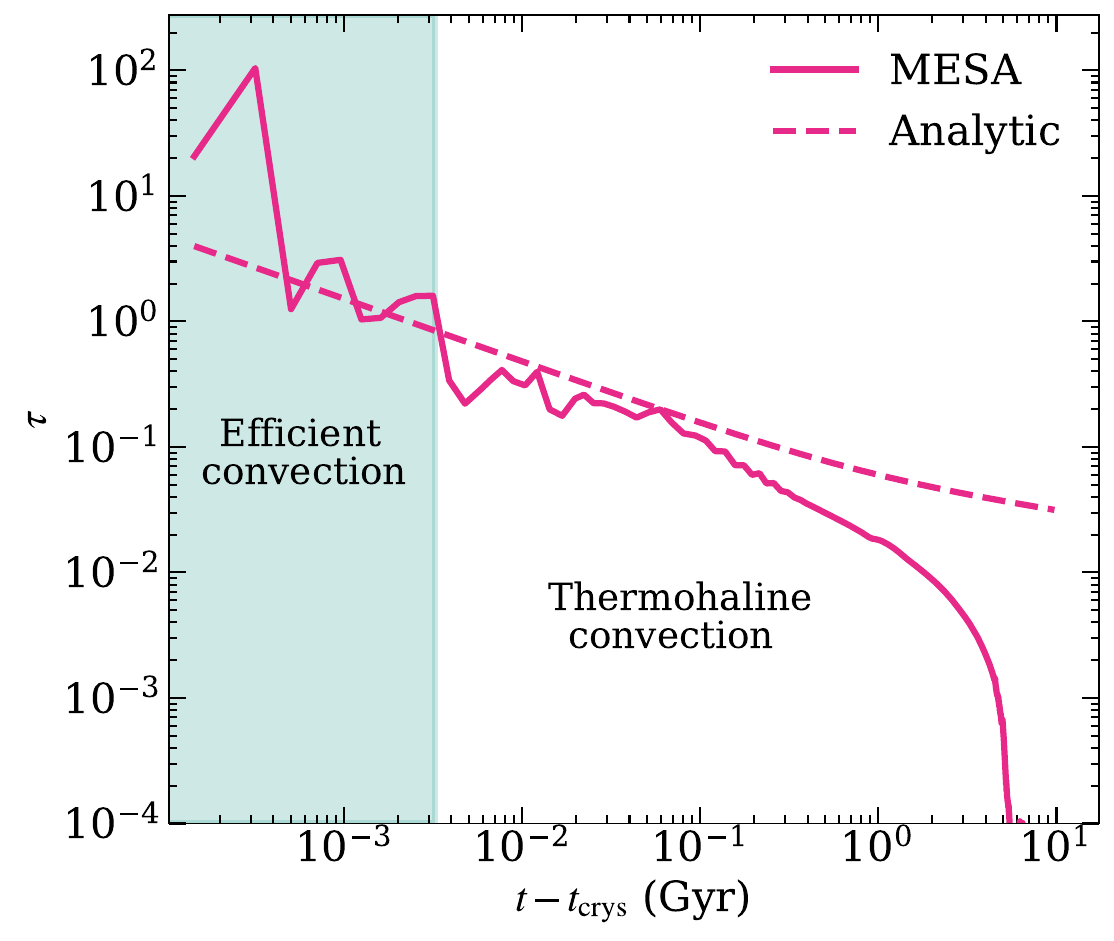}
    \caption{Temporal evolution of the dimensionless composition flux at the solid-liquid interface, $\tau$. Results are shown for a cooling model of a $0.9\msun$ white dwarf obtained with the MESA stellar evolution code. As discussed in \cite{Fuentes2023} and \cite{CT24}, there is a rapid transition between $\tau > 1$, where convection is efficient, and $\tau < 1$, the regime of inefficient thermohaline convection. Note that the small fluctuations are numerical artefacts due to the chemical separation which happens in discrete events in the code. For comparison, we include the analytic prediction for $\tau$ using Mestel's law for cooling white dwarfs (Equation~\ref{eq:tau_analytic}, with $T_i = 7\times 10^6~\mathrm{K}$, and $\alpha = -0.1~\mathrm{Gyr}$)}.
    \label{fig:tau_evolution}
\end{figure}

There are two regimes of convection with a rapid transition between them as $\tau$ increases. For $\tau < 1$, the system undergoes thermohaline (inefficient) convection, while for $\tau > 1$, the fluid undergoes efficient overturning convection. Figure~\ref{fig:tau_evolution} shows that $\tau \sim 1$--$10^2$ at the onset of crystallization, and it remains above unity for $\sim 10~\mathrm{Myr}$. Then, it quickly drops to $\tau \ll 1$ and convection transitions to the thermohaline regime, dominating most of the star's cooling history ($\sim 10~\mathrm{Gyr}$). 

We make an analytic model to confirm that the divergence of $\tau$ towards large values at the onset of crystallization is real and not numerical. Assuming $m(r) \approx (4\pi/3) \rho_c r^3$, where $\rho_c$ is the central density, and combining hydrostatic balance with the equation of state for degenerate electrons $P=K\rho^{5/3}$, where $K$ is a constant, we obtain the density profile $\rho(r) = \rho_c(1-r^2/R^2_{\rm WD})^{3/2}$. The star's interior begins to crystallize when the plasma coupling parameter $\Gamma\propto \rho^{1/3}/T$ (the ratio of the electrostatic to thermal energy) exceeds a critical value $\Gamma_{\mathrm{crit}} \approx 175$ \citep{vanHorn1968,Potekhin2000,Bauer2020,Jermyn2021}. In particular, assuming that $\Gamma_{\mathrm{crit}}$ remains constant during the crystallization, we can write $\rho/\rho_c = (T/T_i)^3$, where $T_i$ is the initial temperature at which the central region of the star begins to crystallize (the external layers of the white dwarf’s have a lower density, and therefore reach $\Gamma_{\mathrm{crit}}$ and crystallize at a later time). Then, we can identify the radius of the solid core as a function of temperature

\begin{equation}
\dfrac{R_{\mathrm{core}}}{R_{\rm WD}} = \left[1 - \left(\dfrac{T_{\mathrm{core}}}{T_i}\right)^2\right]^{1/2}~, \label{eq:Rcore/R}
\end{equation}
where $T_{\mathrm{core}}$ is the core temperature.

Then for an incompressible core $M_{\mathrm{core}} = (4\pi/3) \rho_c R^3_{\mathrm{core}}$, we write $\dot{M}_{\mathrm{core}} = 4\pi \rho_c R^2_{\mathrm{core}}\dot{R}_{\mathrm{core}}$. Taking the time derivative of Equation~\eqref{eq:Rcore/R}, and using $F_X = \dot{M}_{\mathrm{core}}\Delta X/4\pi R^2_{\mathrm{core}}$ (where $\Delta X$ is the carbon enhancement in the liquid phase relative to the solid), Equation~\ref{eq:tau} can be written as

\begin{equation} \label{eq:tau_analytic}
\tau = \alpha \left(\dfrac{\dot{T}_{\mathrm{core}}}{T_i}\right)  \dfrac{(T_i/T_{\mathrm{core}})^3}{\sqrt{1-(T_{\mathrm{core}}/T_i)^2}}~,
\end{equation}
where $\alpha = -R_{\mathrm{WD}}(H_P/\kappa_T)(\Delta X/X)(\chi_X/\chi_T\nabla_{\mathrm{ad}})\approx -0.1~\mathrm{Gyr}$ and it does not vary significantly during the early stages of crystallization. We can see that in this simple model, $\tau$ is mostly determined by the cooling history of the star's core. Adopting Mestel's cooling law $T_{\mathrm{core}} \propto t^{-2/5}$  \citep{Mestel1952}, we find a reasonable agreement with the output from the MESA model (see dashed curve in Figure~\ref{fig:tau_evolution}).

As we previously showed in \cite{Fuentes2023}, $\tau$ can be related to the P\'eclet number Pe (with $\rm Pe\sim 1$ separating the two efficiency regimes). Therefore, we must calculate Pe and the convective velocities for both cases (thermohaline and overturning convection) in order to assess the plausibility of crystallization-driven dynamos.

\section{Convective velocities and stellar dynamo} \label{sec:dynamo}

In this section, we combine the MAC balance and scalings for convection in rotating-magnetized flows to derive analytic expressions for the convective velocity during crystallization, and the strength of the induced magnetic field.

Once the dynamo saturates, convective flows are set by a balance between the magnetic, buoyancy, and Coriolis force (MAC balance)
\begin{equation}\label{eq:MAC}
    {u^2_A\over \ell_{\perp}^2}\sim {g\over \ell_{\perp}}\dfrac{D\rho}{\rho} \sim {2\Omega u_c\over H_P}~,
\end{equation}
where $u_A \equiv B/\sqrt{4\pi \rho}$ is the Alfv\'en speed, $\ell_{\perp}$ is the length scale perpendicular to rotation, and $D\rho/\rho$ is the density contrast in the convective zone. In terms of the composition gradients 
\begin{equation}\label{eq:Drhograd}
{D\rho\over\rho} \approx -{\ell\over 2H_P}{\chi_X\over \chi_\rho}\left(\nabla_X-\nabla_{X,\mathrm{crit}}\right)~,
\end{equation} 
where $\chi_\rho = \left.\partial\ln P/\partial\ln \rho\right|_{T, X}$, $\nabla_X = d\ln X/d\ln P|_\star$ is the light element gradient in the star, $\ell$ is the mixing length, and the critical composition gradient is
\begin{equation}\label{eq:DXcritdef}
\nabla_{X,\mathrm{crit}} \equiv {\chi_T\over \chi_X}\left(\nabla_\mathrm{ad}-\nabla\right) \left({\mathrm{Pe}\over C+\mathrm{Pe}}\right),
\end{equation}
where $C$ is a constant that depends on the shape of a fluid element \citep[e.g., $C=9/2$ in][]{KippenhahnWiegert}. In the limit of large $\mathrm{Pe}$, $\nabla_{X,\mathrm{crit}}$ is the gradient needed to be unstable to overturning convection according to the Ledoux criterion. At small $\mathrm{Pe}$, thermal diffusion reduces the effective thermal stratification, so that a smaller $\nabla_{X,\mathrm{crit}}$ is needed to drive convection. This is the regime of thermohaline (fingering) convection. For more details, see \cite{Fuentes2023}.

\citet{Davidson2013} demonstrated that in flows that have small values of the magnetic Prandtl number, $\mathrm{Pm} \equiv \nu/\eta$ (where $\nu$ is the kinematic viscosity, and $\eta$ is the magnetic diffusivity), the magnetic energy density is determined by the rate of work done by the buoyancy force. This hypothesis and dimensional analysis lead to
\begin{equation}\label{eq:buoyancy production}
u^2_A \sim H_P^{2/3} \left(g \dfrac{D\rho}{\rho} u_c\right)^{2/3}~.
\end{equation}
Typical conditions for the transport properties in the convective region of carbon-oxygen white dwarfs give $\nu\sim 3\times 10^{-2}~\rm cm^2~s^{-1}$, $\eta\sim 6\times 10^{-2}~\rm cm^2~s^{-1}$, resulting in $\rm Pm\sim 0.5$ \citep{Cumming2002,Isern2017}, suggesting that these scalings should apply. Then combining Equations~\eqref{eq:MAC} and \eqref{eq:buoyancy production}, we obtain
\begin{equation}\label{eq:u_A_u_c}
u_c \sim \dfrac{\ell_\perp}{H_P} u_A \sim \left(\dfrac{u^4_A}{2\Omega H_P}\right)^{1/3}~.
\end{equation}
Since the dynamical length scale for convection in rapidly-rotating-magnetized flows is $\ell_{\perp}$, let us use $\ell = \ell_\perp$, so that 
$\mathrm{Pe} = u_c \ell_\perp/\kappa_T$. Rewriting Equation~\eqref{eq:u_A_u_c} as 
\begin{equation}
u_c \ell_\perp \sim \dfrac{u^2_c H_P}{u_A} \sim u^{5/3}_A H_P^{1/3}(2\Omega)^{-2/3}~,
\end{equation}
shows that the Alfv\'en speed is entirely determined by the P\'eclet number, the rotation rate, and the vertical length scale  of the fluid motions
\begin{equation}
u_A \sim (\kappa_T~\mathrm{Pe})^{3/5}  H^{-1/5}_P (2\Omega)^{2/5}~. \label{eq:u_A}
\end{equation}
Equation~\eqref{eq:u_A} can be re-written in terms of the magnetic field $B = u_A\sqrt{4\pi \rho}$
\begin{align}\label{eq:B}
\nonumber B \sim 10^2~\mathrm{G}~\left(\dfrac{\mathrm{Pe}}{1}\right)^{3/5}&\left(\dfrac{\kappa_T}{10~\mathrm{cm^2~s^{-1}}}\right)^{3/5}\left(\dfrac{\rho}{10^7~\mathrm{g~cm^{-3}}}\right)^{1/2}\\
&\times \left(\dfrac{H_P}{10^{8}~\mathrm{cm}}\right)^{-1/5} \left(\dfrac{P_{\mathrm{rot}}}{1~\mathrm{h}}\right)^{-2/5}~,
\end{align}
where $P_{\mathrm{rot}} = 2\pi/\Omega$, and we have used typical values for cooling white dwarfs. Although Equation~\eqref{eq:B} gives a general expression to evaluate the magnetic field as a function of different stellar parameters, we must first calculate the P\'eclet number $\mathrm{Pe}$, which can vary dramatically between the thermohaline regime ($\mathrm{Pe},~\tau \ll 1$) and the efficient regime ($\mathrm{Pe},~\tau \gg 1$). Further, we show later that in the efficient regime, the magnetic field becomes independent of the rotation period $P_\mathrm{rot}$, a fact that is supported by observational data. In what follows, we use mixing-length theory to obtain analytic expressions for Pe for each regime.

\section{Pe for thermohaline and efficient convection} \label{sec:Pe}

\subsection{Regime of small $\mathrm{Pe}$ (thermohaline convection)}

In a mixture of two elements (e.g. carbon and oxygen), the total composition can be described by only one variable, here chosen to be $X$, the mass fraction of the lighter component (carbon in the context of core crystallization). From mixing-length theory, the mass flux of light elements is
\begin{equation}\label{eq:FX}
F_X = \rho u_c DX = \rho u_c X \nabla_X \dfrac{\ell}{2H_P}~,
\end{equation}
where $DX = X \nabla_X (\ell/2H_P)$ is the excess composition carried by a fluid element. Equation \eqref{eq:FX} allows us to write the convective velocity as $u_c \approx (F_X/ \rho X \nabla_X)(2H_P/\ell)$, giving an expression for $\rm Pe$ in terms of the composition flux $F_X$,
\begin{equation}\label{eq:evalPe}
    \mathrm{Pe} \equiv {u_c \ell\over \kappa_T} \approx \left({H_P^2\over \kappa_T}\right) \left({2F_X\over \rho H_P X}\right)\nabla_X^{-1} \approx {t_{\rm therm}\over t_X}{2\over \nabla_X}~.
\end{equation}

In the limit of $\mathrm{Pe} \ll 1$, thermal diffusion reduces the effective thermal stratification, i.e., $\nabla \ll \nabla_{\rm ad}$, where $\nabla = \left.d\ln T/d\ln P\right|_\star$ is the temperature gradient in the star. As a consequence, the critical composition gradient becomes smaller, as we can see by taking the small $\mathrm{Pe}$ limit of Equation \eqref{eq:DXcritdef},
\begin{equation}\label{eq:DXcrit}
\nabla_{X,\mathrm{crit}} \approx\dfrac{\chi_T}{\chi_X} \nabla_{\rm ad}\dfrac{\mathrm{Pe}}{C}.
\end{equation}
The reduction in $\nabla_{X,\mathrm{crit}}$ allows for composition transport with a composition gradient very close to the critical value, $\nabla_X \approx \nabla_{X,\mathrm{crit}}$ \citep[see Section~2.1 in][]{Fuentes2023}. Then, Equation~\eqref{eq:evalPe} becomes
\begin{equation}\label{eq:low_Pe}
	\mathrm{Pe}\approx 3\left({t_{\rm therm}\over t_X}\right)^{1/2}\left({\chi_X \over \chi_T \nabla_{\mathrm{ad}}}\right)^{1/2} \approx 3\tau^{1/2}~,
\end{equation}
where we used $\tau = (t_{\mathrm{therm}}/t_X)(\chi_X/\chi_T\nabla_{\mathrm{ad}})$ and $C=9/2$. Typical convection parameters above the crystallization front for a $0.9~M_{\odot}$ white dwarf give $\tau \sim 10^{-2}$ (see Figure~\ref{fig:tau_evolution}), then $\rm Pe\approx 0.3$, in good agreement with the estimates of \cite{Mochkovitch1983} and \cite{Isern1997}. 

\subsection{Regime of large $\mathrm{Pe}$ (efficient convection)}\label{sec:large_Pe}

Convection in the large $\mathrm{Pe}$ regime is different in that the composition gradient $\nabla_{X}$ significantly exceeds $\nabla_{X,\mathrm{crit}}$ once $\tau>1$ \citep{Fuentes2023}. To find an expression for $\nabla_X$, we can balance the Coriolis and buoyancy forces in Equation~\eqref{eq:MAC}, and rewrite $D\rho/\rho$ in terms of the gradients using Equation~\eqref{eq:Drhograd}. This gives 
\begin{equation}
\dfrac{2\Omega u_c}{H_P} \sim \dfrac{g}{\ell_\perp} {\ell\over 2H_P}{\chi_X\over \chi_\rho}\left(\nabla_X-\nabla_{X,\mathrm{crit}}\right)~.
\end{equation}
Using $\ell = \ell_\perp$, and conveniently introducing $\chi_T$, $\nabla_{\mathrm{ad}}$, and $\kappa_T$, this can be rewritten as
\begin{equation}\label{eq:steps3}
\dfrac{2H_P}{\ell_\perp}\left(\dfrac{\mathrm{Ta}^{1/2}}{\mathrm{Ra_T}}\right)\left(\dfrac{\chi_T \nabla_{\mathrm{ad}}}{\chi_X}\right)\mathrm{Pe}\sim \nabla_X-\nabla_{X,\mathrm{crit}}~,
\end{equation}
where

\begin{align}
&\mathrm{Ta}\equiv \dfrac{4\Omega^2 H_P^4}{\kappa_T^2}\sim 10^{25}\left(\dfrac{P_{\mathrm{rot}}}{1~\mathrm{h}}\right)^{-2}\left(\dfrac{H_P}{10^8~\mathrm{cm}}\right)^4 \left(\dfrac{\kappa_T}{10~\mathrm{cm^2~s^{-1}}}\right)^{-2}~,\\
\nonumber&\mathrm{Ra_T} \equiv \dfrac{gH_P^3\chi_T\nabla_\mathrm{ad}}{\chi_\rho\kappa_T^2}\sim 10^{26} \left(\dfrac{g}{10^8~\mathrm{cm~s^{-2}}}\right)\left(\dfrac{H_P}{10^8~\mathrm{cm}}\right)^3\\
&\hspace{2.7cm}\times \left(\dfrac{\kappa_T}{10~\mathrm{cm^2~s^{-1}}}\right)^{-2} \left(\dfrac{\chi_T\nabla_{\mathrm{ad}}/\chi_{\rho}}{10^{-4}}\right),
\end{align}
are dimensionless parameters that resemble the Taylor and Rayleigh numbers that arise in thermal convection.

To eliminate $\ell_\perp$, we can use Equation~\eqref{eq:u_A_u_c} and write
\begin{equation}\label{eq:L}
\dfrac{\ell_\perp}{H_P} \sim \left(\dfrac{u_A}{2\Omega H_P}\right)^{1/3} \sim \left(\dfrac{u_c}{2\Omega H_P}\right)^{1/4} = \mathrm{Ro}^{1/4}~,
\end{equation}
where $\mathrm{Ro} = u_c/2\Omega H_P$ is the Rossby number. Finally, writing the Rossby number in terms of Pe
\begin{equation}\label{eq:Ro}
\mathrm{Ro} = \mathrm{Pe}^{4/5} \mathrm{Ta}^{-2/5}~,
\end{equation}
and combining Equations~\eqref{eq:steps3}--\eqref{eq:Ro} gives 
\begin{align}\label{eq:nablaX_large_Pe}
&\nabla_X - \nabla_{X,\rm crit} \approx 2 \left(\dfrac{\chi_T \nabla_{\mathrm{ad}}}{\chi_X}\right)\dfrac{\mathrm{Pe}^{4/5} \mathrm{Ta}^{3/5}}{\mathrm{Ra_T}} ~.
\end{align}
Now assuming\footnote{We verified that $\nabla_X>\nabla_{X,\mathrm{crit}}$ by solving the full set of equations using the heat and composition fluxes from the evolution models in a similar way to \cite{CT24} but including MAC balance. For example, we find that $\nabla_X \approx 6\nabla_{X,\mathrm{crit}}$ for the $0.9\ M_\odot$ model shown in Section \ref{sec:cooling}.} that $\nabla_X$ significantly exceeds $\nabla_{X,\mathrm{crit}}$, inserting Equation~\eqref{eq:nablaX_large_Pe} in \eqref{eq:evalPe} and recognizing $\tau = (t_{\mathrm{therm}}/t_X)(\chi_X/\chi_T\nabla_{\mathrm{ad}})$, we get
\begin{align}\label{eq:large_Pe}
\mathrm{Pe} \approx \tau^{5/9} ~ \mathrm{Ra_T}^{5/9}~\mathrm{Ta}^{-1/3}~.
\end{align}
For the model in Figure~\ref{fig:tau_evolution}, $\tau \sim 10$ (on average) during the efficient regime. Then, using $\mathrm{Ra_T} \sim 10^{26}$, and $\mathrm{Ta}\sim 10^{25}$, Equation~\eqref{eq:large_Pe} gives $\mathrm{Pe}\sim 10^6$, approximately seven orders of magnitude larger than in the thermohaline regime. 

\section{Magnetic dynamo at the onset of crystallization}\label{sec:results}

We now evaluate the magnetic field for a set of cooling models of carbon-oxygen white dwarfs with masses $M_{\mathrm{WD}}\approx 0.4$--$1.3 \msun$, calculated with MESA\footnote{The inlists used in this work are publicly available at \dataset[doi:10.5281/zenodo.10622191]{https://doi.org/10.5281/zenodo.10622191}.} as described in Section~\ref{sec:cooling}. These models were constructed by rescaling a $0.6~M_{\odot}$ WD model to the new mass, keeping the same composition profile as a function of relative mass coordinate. We extract the thermodynamic and transport properties above the crystallization front directly from the simulations (see Table~\ref{table:smodel}).  For all our models, the thermohaline regime has $\tau,~\mathrm{Pe} \ll 1$ (e.g.~$\tau \sim 10^{-2}$, $\mathrm{Pe}\approx 0.3$ for the 0.9$\msun$ model, see Section~\ref{sec:Pe}), giving $u_c \lesssim 10^{-5}~\mathrm{cm~s^{-1}}$ and $B \lesssim 100~\mathrm{G}$ for $\mathrm{Pe} \lesssim 1$. This shows that thermohaline convection can not explain observed magnetic fields of $B_{\mathrm{obs}} \sim 10^{4}$--$10^{9}~\mathrm{G}$. We therefore focus on the regime of efficient convection at the onset of crystallization.

\begin{deluxetable*}{lcccccccccccch}
\tablecaption{Thermodynamic and buoyancy properties above the crystallization front for our models of cooling white dwarfs, in the regime of efficient convection. To optimize the construction of the models' grid, with a resolution of $0.1~M_{\odot}$, we use a mass relaxation control in MESA to adjust the WD mass of the settled model in the \texttt{wd\_cool\_0.6M} test suite. As the fast convection occurs on the onset of the crystallization, we increase the time resolution of the simulations before the solid phase appeared, so that the parameters were more accurate at the age $t_{\mathrm{crys}}$, when the crystallization started. For the $1.3~\msun$ model, the outer boundary condition was changed from \texttt{atm\_table='WD\_tau\_25'} to \texttt{'photosphere'} to help convergence.\label{table:smodel}}
\tablewidth{\textwidth}
\tablehead{
\colhead{$M_{\mathrm{WD}}$} & \colhead{$\rho$} & \colhead{$T$} & \colhead{$H_P$} & \colhead{$g$} & \colhead{$\kappa_T$} & $t_{\mathrm{crys}}$ & $\log{(L/L_{\odot})}$ & \colhead{$\chi_T$} & \colhead{$\chi_\rho$} &
\colhead{$\nabla_{\mathrm{ad}}$} & \colhead{$\tau$} & \colhead{$\sigma_\tau$} \\
\colhead{$(\msun)$} &  \colhead{$(\rm g~cm^{-3})$} & \colhead{$(\rm K)$} & \colhead{$\left(\mathrm{cm}\right)$} & \colhead{$(\rm cm~s^{-2})$} & \colhead{$(\rm cm^2~s^{-1})$} & \colhead{(Gyr)} & \colhead{} & \colhead{}  &
\colhead{} & \colhead{} & \colhead{}
}
\startdata
0.4 & $8.53 \times 10^5$ & $2.64 \times 10^{6}$ & $2.91\times10^{8}$ & $7.94\times 10^{7}$ & $5.55\times 10^{1}$ & $2.99$  & $-4.02$ & $7.47 \times 10^{-4}$ & $1.58$ & $0.37$ & 3.10 & 2.65 \\
0.5 & $1.62\times10^6$ & $3.22 \times 10^{6}$ &  $2.72\times10^{8}$ & $1.22\times 10^{8}$ & $6.10\times 10^{1}$ & $2.31$ & $-3.87$ & $6.23 \times 10^{-4}$ & $1.55$ & $0.38$ & 6.46  & 4.51 \\
0.6 & $2.86\times 10^6$ & $3.86\times 10^{6}$ &  $2.47\times10^{8}$ & $1.82\times 10^{8}$ & $6.37\times 10^{1}$ & $1.73$ & $-3.60$ & $5.43 \times 10^{-4}$ & $1.52$ & $0.38$ & 7.56 & 2.48 \\
0.7 & $4.94 \times 10^6$ & $4.63\times 10^{6}$ & $2.22\times10^{8}$ & $2.66\times 10^{8}$ & $6.45\times 10^{1}$ & $1.29$ & $-3.34$ & $4.87 \times 10^{-4}$ & $1.49$ & $0.39$ & 6.76 & 4.66 \\
0.8 & $8.51 \times 10^6$ & $5.57 \times 10^{6}$ & $1.98\times10^{8}$ & $3.88\times 10^{8}$ & $6.31\times 10^{1}$ & $0.96$ & $-3.07$ & $4.47 \times 10^{-4}$ & $1.46$ & $0.39$ & 5.29 & 3.92 \\
0.9 & $1.49 \times 10^7$ & $6.78 \times 10^{6}$ &  $1.71\times10^{8}$ & $5.66\times 10^{8}$ & $6.03\times 10^{1}$ & $0.71$ & $-2.81$ & $4.17 \times 10^{-4}$ & $1.43$ & $0.39$ & 6.72 & 6.02 \\
1.0 & $2.72 \times 10^7$ & $8.39 \times 10^{6}$ & $1.44\times10^{8}$ & $8.83\times 10^{8}$ & $5.52\times 10^{1}$ & $0.51$ & $-2.54$ & $3.93 \times 10^{-4}$ & $1.41$ & $0.40$ & 4.74 & 4.23 \\
1.1 & $5.41 \times 10^7$ & $1.07\times 10^{7}$ & $1.16\times10^{8}$ & $1.44\times 10^{9}$ & $4.85\times 10^{1}$ & $0.35$ & $-2.23$ & $3.74 \times 10^{-4}$ & $1.39$ & $0.40$ & 8.60 & 5.51 \\
1.2 & $1.27 \times 10^8$ & $1.45 \times 10^{7}$ & $8.73\times10^{7}$ & $2.64\times 10^{9}$ & $3.96\times 10^{1}$ & $0.22$ & $-1.88$ & $3.54 \times 10^{-4}$ & $1.37$ & $0.40$ & 38.20 & 26.3 \\
1.3 & $4.41 \times 10^8$ & $2.22 \times 10^{7}$ &  $5.68\times10^{7}$ & $6.32\times 10^{9}$ & $1.92\times 10^{1}$ & $0.12$ & $-1.41$ & $4.51 \times 10^{-4}$ & $1.35$ & $0.35$ & 14.30 & 7.58 \\
\enddata
\end{deluxetable*}

\begin{figure}
    \includegraphics[width=\columnwidth]{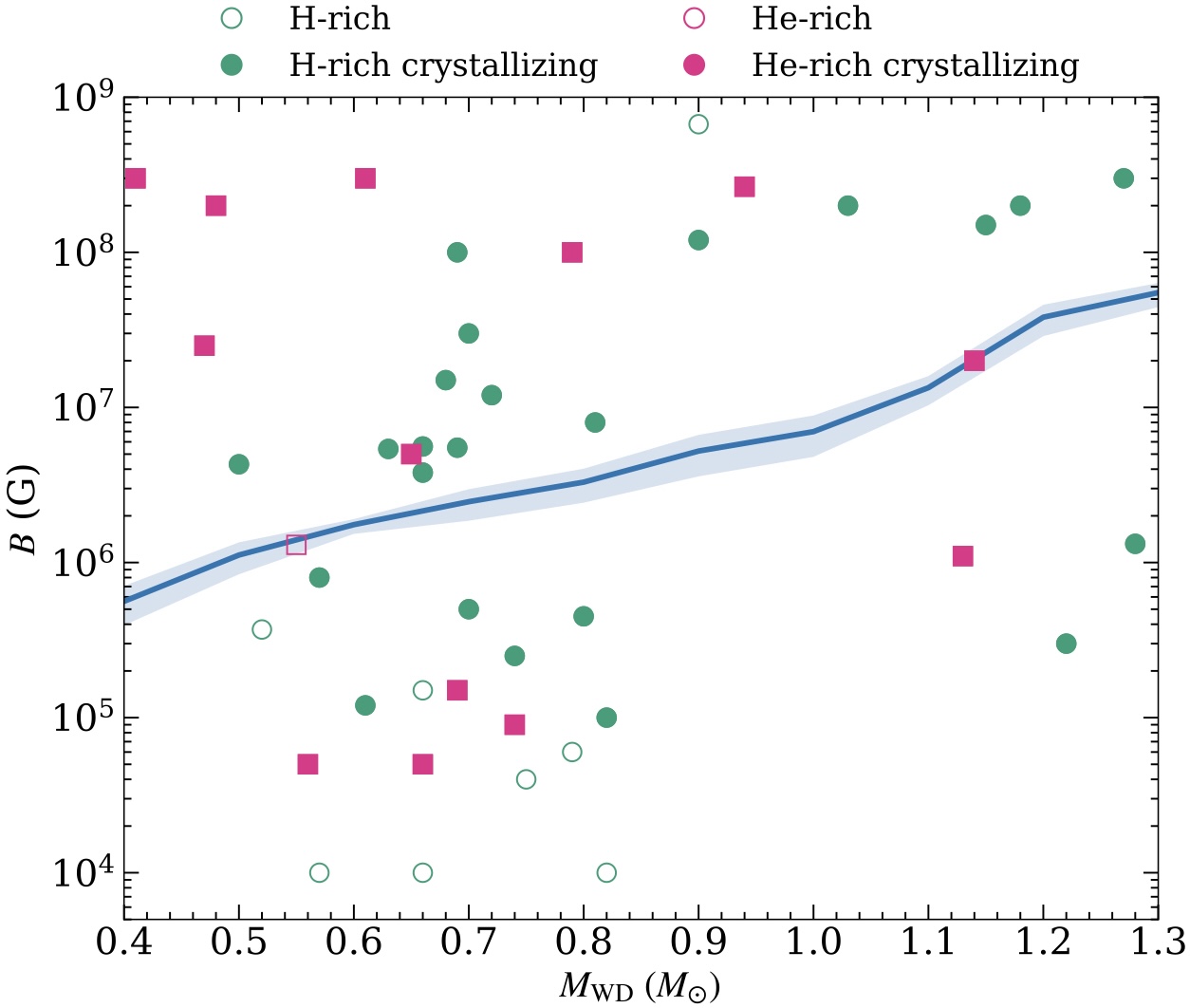}
    \caption{Intensity of the magnetic field generated at the onset of crystallization, as a function of the white dwarf’s mass. The solid line corresponds to predictions from Equation~\eqref{eq:B2}, where the physical parameters above the crystallization front were extracted from a grid of MESA models. The shaded region corresponds to the uncertainties in the predictions due to the dispersion in the measurements of the dimensionless composition flux $\tau$ at the early phase of crystallization. We include measurements of observed magnetic fields in isolated white dwarfs taken from \citet{Bagnulo2022}. We classified the observational data in 4 different groups (H and He rich, and whether they are crystallizing or not based on the age at which the MESA models begin crystallization). }
    \label{fig:B_vs_M}
\end{figure}

It is useful to combine Equations~\eqref{eq:B} and \eqref{eq:large_Pe} to obtain an expression for the magnetic field in the limit of efficient convection in terms of the local stellar properties,
\begin{align}\label{eq:B2}
\nonumber B \approx 1.4~\mathrm{MG}~\left(\dfrac{\tau}{10}\right)^{1/3} &\left(\dfrac{\rho}{10^7~\mathrm{g~cm^{-3}}}\right)^{1/2} \left(\dfrac{g}{10^8~\mathrm{cm~s^{-2}}}\right)^{1/3}\\
&\times\left(\dfrac{\kappa_T}{10~\mathrm{cm^2~s^{-1}}}\right)^{1/3} \left(\dfrac{\chi_T\nabla_{\mathrm{ad}}/\chi_{\rho}}{10^{-4}}\right)^{1/3}~.
\end{align}
Note that this is independent of the star's rotation rate. Figure~\ref{fig:B_vs_M} shows the predicted magnetic field strength as a function of white dwarf mass. We see that our calculation for crystallization-driven dynamos using the MAC balance can naturally explain many of the observed values of magnetic field, with the predicted values lying in the range  $B\sim 10^6$--$10^8\ \mathrm{G}$. The predicted values correlate strongly with white dwarf mass, whereas the observed magnetic fields do not. 
It is interesting to note that when separated into H-rich and He-rich, the hydrogen ones seem to follow qualitatively a similar trend to the model (although with large dispersion). 
Note however that a comparison between the predicted and observed magnetic fields requires modelling the transport of magnetic field from the core where it is generated to the surface \citep{Blatman2024}.

Finally, for the cooling models investigated in this work, Equations~\eqref{eq:u_A_u_c}, \eqref{eq:u_A}, and \eqref{eq:large_Pe},   give $u_c\sim 10^2$--$10^3~\mathrm{cm~s^{-1}}$ in the efficient regime, increasing with the mass of the star. 
Comparing the kinetic and magnetic energy densities, we obtain $B^2/8\pi \sim 10^{10}$--$10^{14}~\mathrm{erg~cm^{-3}}$ and $\frac{1}{2}\rho u^2_c\sim 10^9$--$10^{14}~\mathrm{erg~cm^{-3}}$, meaning that equipartition is approximately satisfied in the regime of efficient convection.

\section{Discussion}\label{sec:discussion}

Our results suggest that the crystallization-driven dynamo operates in a short intense phase following the onset of crystallization, 
before convection transitions to the thermohaline regime. The long Ohmic diffusion timescales in white dwarfs, particularly once the fluid solidifies (e.g.~\citealt{Cumming2002}) means that we can expect the strong field created in these initial stages of crystallization to survive for the remaining lifetime of the white dwarf.

Unlike previous works that rely on energy arguments \citep{Isern2017} or the CIA balance \citep[e.g.,][]{Ginzburg2022,Fuentes2023}, our estimations of the convective speed and magnetic fields take into account the correct force balance (Lorentz, Coriolis, and buoyancy force). Our analytic expression for the magnetic field is independent of the rotation rate of the star (see Equation~\ref{eq:B2}), a fact that is supported by observations \citep[e.g.,][]{Kawka2020,Ferrario2020}. For the set of white dwarf cooling models we have used here, the predicted field strength increases with mass (Figure \ref{fig:B_vs_M}) which is not seen in the data, although more work is needed to investigate a more complete set of white dwarf models and to calculate the evolution of the field through the phase of thermohaline convection. The fact that the convection is driven from a small central solid core during the efficient dynamo phase is likely to result in a large angular scale, ie.~dipole-dominated, magnetic field geometry.

As pointed out recently by \cite{Blatman2024}, another important problem for crystallization-driven dynamos in white dwarfs is emergence of the magnetic field at the star's surface. Initially, the field is trapped in the convection zone, deep inside the core. Only later on in the evolution of the star, when the convection zone has expanded outward closer to the surface, is the magnetic diffusion time \citep[e.g.,][]{Cumming2002,Saumon2022} short enough for the field to break out to the surface and be observed. The extent of the convection zone during the efficient convection phase depends on the initial carbon-oxygen abundance profile which is set during stellar evolution and depends on the star's mass \citep[with the convective mantle of massive white dwarfs being closer to the surface, e.g.,][]{Bauer2023}. Therefore, the observed strength of the magnetic field will depend mainly on 1) transport of magnetic field between the convection zone and the surface, and 2) how thermohaline convection driven at later times during the cooling history of the star responds to the large magnetic field induced at the onset of crystallization. For this, simulations of magnetized-thermohaline convection, similar to the ones carried out by \cite{Harrington_Garaud_2019} and \cite{Fraser2023} will be particularly useful.\vspace{-0.14in}

\begin{acknowledgements}
We thank James Fuller and Sivan Ginzburg for useful comments on an early version of this manuscript. J.R.F. is supported by NASA through grants 80NSSC19K0267 and 80NSSC20K0193. A.C. and M.C.-T. acknowledge support by NSERC Discovery Grant RGPIN-2023-03620. A.C. and M.C.-T. are members of the Centre de Recherche en Astrophysique du Québec (CRAQ) and the Institut Trottier de recherche sur les exoplanètes (iREx).
\end{acknowledgements}\vspace{-0.125in}

\bibliography{references}{}
\bibliographystyle{aasjournal}

\end{document}